\documentstyle[11pt,epsfig,aaspp4]{article}

\singlespace

 1

\def\be{\begin{equation}}
\def\ee{\end{equation}}

\def\etal{{\it et al.~}}

\def\gs{\mathrel{\raise1.16pt\hbox{$>$}\kern-7.0pt
\lower3.06pt\hbox{{$\scriptstyle \sim$}}}}
\def\ls{\mathrel{\raise1.16pt\hbox{$<$}\kern-7.0pt
\lower3.06pt\hbox{{$\scriptstyle \sim$}}}}
\def\gtsima{$\; \buildrel > \over \sim \;$}
\def\ltsima{$\; \buildrel < \over \sim \;$}
\def\prosima{$\; \buildrel \propto \over \sim \;$}
\def\gsim{\lower.5ex\hbox{\gtsima}}
\def\lsim{\lower.5ex\hbox{\ltsima}}
\def\simgt{\lower.5ex\hbox{\gtsima}}
\def\simlt{\lower.5ex\hbox{\ltsima}}
\def\simpr{\lower.5ex\hbox{\prosima}}

\def\pp{\noindent\parshape 2 0truecm 17truecm 2truecm 15truecm}
\def\rf#1;#2;#3;#4 {\par\pp#1, #2, #3, #4. \par}

\def\pr{\ref@jnl{Phys.Rev}}

\def\href#1;#2 {{\bf #1} : {\em #2}}

\def\beq#1{\begin{equation}\label{#1}}
\def\eeq{\end{equation}}
\def\beqa#1{\begin{eqnarray}\label{#1}}
\def\eeqa{\end{eqnarray}}

\def\H2p{H$_2^+$ }

\def\mH2p{H_2^+}



\begin{document}

\title{CMB ANISOTROPIES DUE TO FEEDBACK-REGULATED INHOMOGENEOUS REIONIZATION}

\author{Marialuce Bruscoli\altaffilmark{1}, Andrea
Ferrara\altaffilmark{2}, Roberto
Fabbri\altaffilmark{3}, Benedetta Ciardi\altaffilmark{2}}

\begin{abstract}
We calculate the secondary anisotropies in the CMB produced by inhomogeneous
reionization from simulations in which the effects of radiative and stellar
feedback effects on galaxy formation have been included. This allows to
self-consistently determine the beginning ($z_i\approx 30$), the duration ($
\delta z\approx 20$) and the (nonlinear) evolution 
of the reionization process for a
critical density CDM model. In addition, from the simulated spatial
distribution of ionized regions, we are able to calculate the evolution of
the two-point ionization correlation function, $C_\chi$, and obtain the
power spectrum of the anisotropies, $C_\ell$, in the range $5000 < \ell <
10^6$. The power spectrum has a broad maximum around $\ell
\approx 30000$, where it reaches the value $2\times 10^{-12}$. 
We also show that the angular correlation function $C(\theta )$ 
is not Gaussian, but  at separation angles $%
\theta \lower.5ex\hbox{\ltsima} 10^{-4}$~rad it can be approximated
by a modified Lorentzian shape; at larger separations an
anticorrelation signal is predicted.
Detection of signals as above
will be possible with future mm-wavelength interferometers like ALMA,
which appears as an optimum instrument to search for signatures of
inhomogeneous reionization.
\end{abstract}


\thispagestyle{empty}

\altaffiltext{1}{Dipartimento di Astronomia, Universit\`a degli studi 
di Firenze, L.go E. Fermi 5, Firenze, Italy} 
\altaffiltext{2}{Osservatorio Astrofisico di Arcetri, L.go E. Fermi 5,
Firenze, Italy} 
\altaffiltext{3}
{Dipartimento di Fisica, Universit\'a di Firenze, Via S. Marta 3, I-50139 
Firenze, Italy}



\section{INTRODUCTION}
 
At $z\approx 1100$ the intergalactic medium recombined and remained neutral
until the first sources of ionizing radiation form and begin 
to reionize it. Current
models of cosmic structure formation predict that the first collapsed,
luminous objects should have formed at redshift $z\approx 30$. This
conclusion is reached by requiring that the cooling time $t_c$ of the gas
be shorter than the Hubble time $t_H$ at the formation epoch. The ionizing
flux from these objects creates cosmological HII regions in the surrounding
IGM, whose sizes are much smaller than their typical interdistance (Ciardi,
Ferrara \& Abel 1999). This implies that reionization passed through a
highly inhomogeneous phase which ended only when the individual HII regions
overlapped, or stated differently, when reionization was complete.
There is essentially no direct observational test of both the beginning and the
duration of the reionization process.  The most stringent upper limits 
on the reionization redshift and the Thomson optical depth 
of the cosmic medium, derived from data on ``linear'' anisotropies at $%
\ell < 10^3$, are quite
model-dependent (De Bernardis \etal 1997, Griffiths \etal 1999). 

However, if reionization was indeed inhomogeneous, it should have left an
inprint in the CMB. In the homogeneous case, small-scale
secondary anisotropies generated by Doppler effect from
Thomson scattering off IGM electrons would be erased due to
potential flow cancellation effects. In the
inhomogeneous case, the modulation of the ionization fraction, playing a
similar role as the density modulation for the nonlinear Vishniac effect,
prevents such cancellation leading to anisotropies at sub-degree scales.
A few papers have recently tackled the
calculation of secondary anisotropies produced by inhomogeneous reionization
(Aghanim {\it et al.~} 1996 [see also Erratum (1999)], Knox, Scoccimarro \&
Dodelson 1998, Gruzinov \& Hu 1998, Peebles \& Juszkiewicz 1998, 
Haiman \&
Knox 1999, Hu 1999). These models are based on simple assumptions concerning
the evolution of the mean ionization level, the typical size of the ionized
patches and, in some cases, their spatial and redshift correlation.

Our aim here is to improve the modelling by self-consistently calculating
secondary anisotropies from dedicated reionization simulations presented in
Ciardi, Ferrara, Governato \& Jenkins (1999, CFGJ), in which the effects of
radiative and stellar feedback effects on galaxy formation have been
included. Similar reionization simulations have been performed also by
Gnedin (1998, 1999). This allows us to calculate the history and correlation
properties of reionization to an unprecedented detail level.

\section{INHOMOGENEOUS REIONIZATION}

The predictions for the CMB anisotropies presented in this paper are based
on the study of inhomogeneous reionization (IHR) presented by CFGJ. Although
the detailed description of the model can be found in that paper, it is
useful to recall here its essential features and the results relevant for
the present work.

The reionization process is studied in a critical density CDM universe ($%
\Omega_0=1$, $h=0.5$ with $\sigma_8=0.6$ at $z=0$ and $\Omega_b=0.06$) as
due to stellar sources, including Population III objects. The spatial
distribution of the sources is obtained from high-resolution numerical
N-body simulations within a periodic box of comoving length $L=2.55h^{-1}$
Mpc. The source properties are calculated taking into account a
self-consistent treatment of both radiative ({\frenchspacing{\it i.e. }}
ionizing and H$_2$--photodissociating photons) and stellar ({\frenchspacing%
{\it i.e. }} SN explosions) feedbacks regulated by massive stars. This
allows, in particular, to derive the spatial distribution of the ionized
regions at various cosmic epochs and the evolution of the mean H ionization.
In brief, there are two main free parameters in the simulations: ({\it i})
the fraction of total baryons converted into stars $f_{b\star}$, and ({\it %
ii}) the escape fraction of ionizing photons, $f_{esc}$, from a given
galaxy; a critical discussion of these parameters is given in CFGJ. Four
different combinations of these parameters in runs A ($f_{b\star}=0.012,
f_{esc}=0.2$), B (0.004, 0.2), C (0.15, 0.2) and D (0.012, 0.1) have been
explored. Run A gives the best agreement between the derived evolution
of the cosmic star formation rate and the experimentally deduced one at 
$z\simlt 4$ (Steidel \etal 1998, CFGJ). 
Therefore, here we only present results for this case.

The topological structure of ionization at $z=19.8$ for run A is shown in
Fig. \ref{fig1}. The proper size of ionized bubbles is approximately in the
range 1 - 20 kpc at this epoch, as only relatively small, and hence faint,
objects have collapsed; they grow in number and volume with cosmic time.%
\footnote{%
Additional simulation images can be found at {\tt %
http://www.arcetri.astro.it/$\sim$ferrara/reion.html/}} A more global view
of the reionization process is given in Fig. \ref{fig2}, where the redshift
evolution of the volume-averaged mean ionization fraction, $\bar\chi(z)$ is
shown for the four different runs. Except for run C (high star formation
efficiency), when reionization is complete at $z\approx 15$, primordial galaxies
are able to reionize the IGM at a redshift $z\approx 10$. Note that
reionization begins at $z_i\approx 30$ when the conditions of the cosmic
medium allow
the first Pop III objects to start to form, and it is then completed after a
redshift interval $\delta z \approx 20$ thanks to the contribution of
larger objects with mass $\approx 10^{8} M_\odot$.

\section{SECONDARY ANISOTROPIES}

To calculate the secondary anisotropies produced by IHR, we use the same
method outlined by Knox {\it et al.}~(1998) and Gruzinov \& Hu (1998). The
solution of the Boltzmann equation for the present value of the perturbation
of the photon temperature $\Delta \equiv \delta T/T$
can be written as 
\begin{equation}
\Delta_0(\hat\gamma) =\tau_0 \int_{0}^1 {\frac{d\eta }{\eta^3}} \chi({\bf x}%
,\eta) \hat\gamma \cdot {\bf v}({\bf x}),
\end{equation}
where $H_0=100 h$~km~s$^{-1}$~Mpc$^{-1}$ is the Hubble constant, $%
\eta=2/H_0(1+z)^{1/2}$ is the conformal time, $\tau_0=n_{e,0}\sigma_T\eta_0
c =0.137 \Omega_b h/\mu$ with $n_{e,0}$ being the present free electron
density, $\sigma_T$ the Thomson cross section, and $\mu=0.59$ the mean
molecular weight of a cosmological mixture of ionized H and He. The quantity 
$\chi({\bf x},\eta)$ is the ionization fraction calculated at position ${\bf %
x}=\hat\gamma(\eta_0-\eta)$ and conformal time $\eta$; ${\bf v}({\bf x})$ is
the peculiar velocity today in units of $c$. The two-point correlation
function due to ionization is then 
\begin{equation}
C(\theta) \equiv \langle \Delta_0(\hat\gamma_1)
\Delta_0(\hat\gamma_2)\rangle \vert_{\cos\theta=\hat\gamma_1\cdot
\hat\gamma_2},  \label{cteta}
\end{equation}
or, using the expression for $\Delta_0(\hat\gamma)$ above, 
\begin{equation}
C(\theta) =\tau_0^2 \int_{0}^1 {\frac{d\eta_1 }{\eta_1^3}} \int_{0}^1 {\frac{%
d\eta_2 }{\eta_2^3}} C_v(\eta_1,\eta_2,\theta) C_\chi(\eta_1,\eta_2,\theta),
\end{equation}
where $C_v(\eta_1,\eta_2,\theta)=\langle\hat\gamma_1 \cdot {\bf v}({\bf x_1}%
) \hat\gamma_2 \cdot {\bf v}({\bf x_2})\rangle$ and $C_\chi(\eta_1,\eta_2,%
\theta)=\langle\chi({\bf x_1},\eta_1) \chi({\bf x_2},\eta_2)\rangle$ (we
have implicitly assumed that the velocity and ionization are independent
fields). The velocity correlation function depends on the matter power
spectrum $P(k)$. Groth {\it et al.~} (1989) have shown that $C_v$ can be
written as 
\begin{equation}
C_v(\eta_1,\eta_2,\theta)=P_{12}\Pi(r)+S_{12}\Sigma(r),
\end{equation}
where $P_{12}=(\hat \gamma_1 \cdot {\bf r})(\hat \gamma_2 \cdot {\bf r})/r$, 
$S_{12}=\hat \gamma_1 \cdot \hat \gamma_2 -P_{12}$ and ${\bf r}= {\bf x_1} - 
{\bf x_2}$. The functions $\Pi(r)$ and $\Sigma(r)$ can be written (for a
vanishing cosmological constant) as 
\begin{equation}
\Pi(r)=(H_0^2\Omega^{1.2}/2\pi^2)\int_0^\infty dk P(k)j_0(kr)-2\Sigma(r),
\end{equation}
\begin{equation}
\Sigma(r)=(H_0^2\Omega^{1.2}/2\pi^2)\int_0^\infty dk P(k)(kr)^{-1}j_1(kr).
\end{equation}
We have adopted a CDM power spectrum of fluctuations as in Efstathiou, Bond
\& White (1992): 
\begin{equation}
P(k)= A k \left\{1+ \left[Bk + (Ck)^{3/2} + (Dk)^{2} \right]^{\nu}
\right\}^{-2/ \nu},  \label{spectrum}
\end{equation}
where $\nu=1.13$, $B=6.4$ $h^{-2}$ Mpc, $C=3.0$ $h^{-2}$ Mpc and $D=1.7$ $h^{-2}
$ Mpc. The power spectrum and its normalization are consistent with the
above numerical simulations.

The most delicate point of the calculation of reionization anisotropies is
probably the form of $C_\chi$. Previous studies (see the Introduction) have
often computed this function assuming a superposition of ionized
patches of equal size R. In addition, the initial redshift of reionization $%
z_i$, its duration $\delta z$, and the evolution of the mean ionization
fraction $\bar\chi(\eta)$ had to be postulated there. Finally, the spatial
correlation of the ionizing sources is also crucial  (for a discussion see
Oh 1998) as it might induce
analogous correlations in the ionization field. In their analytic treatment
Knox \etal 1998 have been forced to postulate an ad-hoc functional 
form for this quantity. Thanks to the IHR
simulations presented above, we are instead able to calculate $C_\chi$ without
further assumptions as explained in the next Section.

\section{ THE IONIZATION CORRELATION FUNCTION}

To calculate $C_\chi$ we proceed as follows. Our simulations provide us with
11 data boxes containing the spatial information on the ionization field for
the following redshifts $z=8.3,10.9,11.4,14.3,15.4,16.5,18.0,19.8,22.1,25.3$, and
$29.6$. As an example, we have shown in Fig. \ref{fig1} the one corresponding to $z= 19.8$.  
We label each epoch with an index 
$n=1,...,9$, where $n=1$ ($n=9$) corresponds to $z=10.9$ ($z=25.3$). For each
box we select a slice through its center, thus obtaining a $N\times N = 256^2$
array with the values of the ionization fraction at each cell point. 
The first and the
last slice have $\bar\chi=1$ (fully ionized) and $\bar\chi=0$ (completely
neutral), respectively. Therefore they do not contribute to 
anisotropies from inhomogeneous reionization and
can be disregarded in the calculation.
For each couple of slices with redshifts $z_n$, $z_m$ we then calculate the
total two-point correlation function as 
\begin{equation}
{\cal C}^{nm}_\chi (\theta, z_n,z_m)=\sum_{i=1}^{N}\sum_{j\ge i}^{N}{\frac{%
\chi^n(i) \chi^m(j)}{N_\theta}}.
\end{equation}
The $(1/2)N(N-1)$ components of $C^{nm}_\chi$ (for each couple $n,m$) thus
obtained are binned according to the separation angle $\theta$ between
the two lines of sight passing through the cells $i$ and $j$.
Hence, $\theta$ runs from $\theta_{min} = 2.0\times 10^{-6}$~rad, the angle
subtended by the cell proper size at the highest redshift of the simulation,
to $\theta_{max}=8.4\times 10^{-4}$~rad, corresponding to proper size $\sqrt{%
2} L$ at the lowest redshift considered; the angular distance used is the
standard Friedmann one. $N_\theta$ is then the number of values falling in $%
[\theta,\theta+\theta_{min}]$. We repeat the procedure for a random slice
realization with the same volume-averaged mean of the ionization, thus
obtaining ${\cal C}_r(\theta)$. Assuming Poisson noise, the associated statistical 
error on the correlation functions is ${\cal C}(\theta)/\sqrt{N_\theta}$.

The correlation function due to ionized patches ({\frenchspacing{\it i.e. }}
the inhomogeneous part) can then be written as: 
\begin{equation}
C^{nm}_\chi(\theta, z_n, z_m)={{{\cal C}^{nm}_\chi(\theta, z_n, z_m)- {\cal C%
}^{nm}_r(\theta, z_n, z_m)}}.
\end{equation}
This function correctly describes reionization effects due to ionized
patches: for example, at the lowest redshift of our simulations, where the
reionization is complete, $C_\chi(\theta)=0$.

In Fig.\ref{fig3} we show the autocorrelation functions $C_\chi^{nn}$ along with the
corresponding errors; the largest error in the curves is $1.9$\%
The behavior of $C_\chi$ is rather flat for small angles ($\theta \lower.5ex%
\hbox{\ltsima} 10^{-5}$~rad) and decreases rapidly for larger separations;
for even larger values of $\theta$, $C_\chi$ becomes negative, implying that
the ionization regions are anticorrelated. The position of the first zero, $%
\theta_0$, provides an estimate of the typical size of the ionized regions
(including overlapping) at the corresponding redshift through the relation $%
R\simeq 0.23 (\theta_0/10^{-4}{\rm rad})h^{-1}$~Mpc. The amplitude of the
anticorrelation signal is roughly proportional to the correlation one.
For small separations $C_\chi^{nm}$ can be approximated by
a modified Lorentzian function  of the type 
\be 
\frac{A_{max}}{1\;+\;[(\theta-\theta_{min})/\Gamma]^{\alpha}}\;,
\label{fit}
\ee
where $A_{max}$ is the maximum amplitude at $\theta=\theta_{min}$, 
and $\Gamma$ is the width, $\alpha$ is the power index.
The values of these fitting parameters for the
autocorrelation functions are given in Tab. 1.

The anticorrelated
regions cannot be described in terms of such a simple assumption. 
Therefore the shape is different from the 
Gaussian one proposed by Gruzinov \& Hu 1998) with
redshift-dependent amplitude and standard deviation. 
The autocorrelation functions for any $n$ are always larger than those obtained
by correlating slices at different redshifts ($m\neq n$), although the shape
remains similar. We have also checked that the results do not depend on the
particular slice used provided it is located not too far from the center
of the simulation box. Off-center planes might lead to sensibly different
amplitudes of ${\cal C}_\chi$ ($\approx \pm 20$\%) due
to border effects, only if their position is closer than $0.04 L$ to 
the box walls.

\section{RESULTS AND CONCLUSIONS}

By using the derived ionization correlation function, we can calculate the
angular power spectrum produced by the inhomogeneous part of the
reionization from eq. \ref{cteta}. In Fig. \ref{fig4}
we show the behavior of the two-point correlation function, $C(\theta)$, and
of the angular power spectrum $C_\ell$ which, for $%
\ell \gg 1$, can be approximated by 
\[
C_\ell = 2\pi \int_0^\infty d\theta \theta J_0(\ell\theta) C(\theta).
\]
The correlation function smoothly decreases from its maximum $C(0)
\approx 5\times 10^{-12}$ and becomes negative at $\theta=1.7\times 10^{-4}$%
~rad as a result of the anticorrelation discussed above. Beyond the second
null point ($\theta \approx 5\times 10^{-4}$~rad) it rapidly becomes more
noisy due the decreasing number of data points in the simulation at these
large separations. In addition to the statistical error, this region is also
affected by systematic errors. These are mainly due to the finite size
of the box which does not allow to take into account ionized regions
produced by sources located just outside the simulation volume.

To minimize the of this effect we have cut the spectrum beyond
separations that are larger than the minimum difference between the angular
size of the box and the typical angular size of the ionized regions $\theta_0
$ (see previous Section). This procedure should yield a $C(\theta)$
reliably unaffected by systematic errors below $\theta_c \approx
6\times 10^{-4}$~rad: we therefore limit our analysis to the angular
interval $\theta_{min} < \theta < \theta_c$.

The reported angular power spectrum therefore extends 
in the range $5000 \lower.5ex%
\hbox{\ltsima} \ell \lower.5ex\hbox{\ltsima} 10^6$ because of the
relation $\ell = \pi / \theta $. The spectrum shows an 
absolute maximum around $%
\ell \approx 30000$, where it has the value $2\times 10^{-12}$. Beyond $\ell
\approx 10^5$ ringings appear in Fig. 4 which are 
caused by the abrupt cutoff of $%
C(\theta)$ at $\theta_c$, but up to  $\ell
\approx 10^5 $ the behaviour of angular spectrum is 
not very far from flatness.
We have calculated a running average $S_\ell$ of the spectrum 
(also shown in Fig. \ref{fig4}) through the formula
\begin{equation}
S_\ell ={1\over 2\pi} \sum_{\lambda =\ell/2}^{3\ell/2}\lambda C_\lambda, 
\end{equation}
This allows to get
rid of the ringings (which have essentially zero average) and to appreciate
the decrease to zero of the spectum above $\ell = 2\times 10^5$.
Our results match very closely those obtained
by Knox {\it et al.~} (1998). In particular,
the best agreement is found with their correlated-source model 
corresponding to $z_i =35,
\delta z=4$. On the other hand, our numerically derived values are $z_i \approx
30, \delta z\approx 15$ (also notice that our evolution of $\bar \chi(z)$ is
nonlinear). Therefore the close agreement can be regarded as rather accidental,
given also
the various assumptions made in that work; we
recall that we have derived here the spatial distribution and evolution of
the ionization fraction from numerical simulations in which several feedback
effects regulating the formation of the ionizing sources have been included.
Also shown for comparison in Fig. \ref{fig4} are the primary angular spectra
corresponding to a CDM model with the same parameters adopted here, in which 
the effects of a homogeneous reionization with total $\tau=0.177$ (as derived from
our simulations) are either neglected or includuded. These spectra have been
obtained by running CMBFAST (Seljak \& Zaldarriaga 1996).
The maximum value of our angular spectrum is comparable to, but higher 
than the most recent
estimates of the Vishniac signal (Jaffe \& Kamionkowski 1998, Hu 1999). For
a cluster-normalized CDM model (which is thereby directly comparable with ours),
Jaffe \& Kamionkowski (1998) find a peak signal $\approx 8\times 10^{-13}$, 
which is however located at lower
multipole numbers, around $\ell \approx 5000$. We also note that our predicted
maximum value is typically smaller than the peak anisotropy
in un uncorrelated-source models with large ionized bubbles.
For instance it is about 3 times smaller than the estimate given by
Gruzinov \& Hu (1998), for their favorite model with $z_i=30, \delta z=5, R=3
$~Mpc [the difference is smaller if we consider more refined 
computation of Knox et al (1999)]. For such uncorrelated-source models
however a large bubble size implies that the peak is confined to
$\ell \simlt 5000$. In any case, all of these models predict that the
signal from IHR should dominate both the primary and the Vishniac one for $%
\ell \lower.5ex\hbox{\gtsima} 5000$.

The detection of this signal would be of great importance for the study of
early galaxy formation, the IGM and the nature of the reionization sources.
If reionization is predominantly produced by stellar type objects, as
assumed here, the relatively small, sub-Mpc sizes of the ionized patches and
their correlation properties, tend to shift the peak of the power spectrum
towards larger multipole values, $\ell \approx 30000$. This detection will
require the use of the next generation of millimiter wavelength
interferometers like ALMA \footnote{{\tt http://www.eso.org:8082/info/}}%
.This instrument is expected to reach sensitivities of 2 $\mu$K in one hour
and reach $\ell \approx 10^6$, thus appearing as a perfect instrument to
search for signatures of IHR, and, in general, for the above type of
studies. Coupled with theoretical predictions for different cosmological
models and simulations exploring a wide range of variation of the key
parameters, as $f_{b\star}$ and $f_{esc}$, these experiments can give a
comprehensive and clean view of how reionization proceeded in the universe.

Instruments nearer in the future as {\it MAP} and {\it %
PLANCK} could at least clarify and constrain the role, if any, played by
quasars in the reionization process. Mini-quasars have been advocated by
some authors (Haiman, Madau \& Loeb 1999) as possible sources for
reionization; in this scenario, we would expect larger patch sizes and a
spectrum peaked at lower values of $\ell$, as in the uncorrelated-source
models with $R$ of order a few Mpc. However, uncertainties might
remain related to the duty-cycle of such objects, if they exist.

IHR might in principle affect the determination of cosmological
parameters planned with future experiments. Our calculations do not allow us
to quantify the uncertainty introduced because the finite size of the
simulation box prevents extension of our results below $\ell \approx 5000$,
where the primary spectrum should be already more than one order of
magnitude smaller as shown by Fig. \ref{fig4}. 
However, since from our results the angular spectrum appears to be
already declining for decreasing $l$ for $l < 10^4$,
a comparison with the results of Knox {\it et al.~} (1998) 
seems to imply that the impact 
of IHR on PLANCK may somewhat less than in their analytical model.
Those authors also conclude that the very small scales
affected by IHR do not sensibly alter the parameter determination obtained
by MAP; our results further support their conclusion. 

Although our calculations do not include the polarization angular 
spectrum $C_{P \ell }$, a rough estimate can be obtained 
extrapolating  the analytic results of Hu (1999) to the present
case. According to this author $C_{P \ell } / C_{ \ell } \approx 10^{-1}%
(Q_{rms}/v_{rms})^2$, where the ratio of rms quadrupole to rms velocity field
is evaluated at $z = z_i$ assuming $\delta z/z_i \ll 1$.  Although the
simulations show that $\delta z/z_i \approx 1$, we can nevertheless
infer that $C_{P \ell } / C_{ \ell } \approx 10^{-5}$. Since for $\tau = 0.177$
the  ``linear'' angular spectrum has $C_{P \ell } / C_{ \ell } \approx %
10^{-2}$ for large $\ell $,
we conclude that the impact of IHR on cosmological parameter 
determination by PLANCK will be even less when polarization data
are taken into account. The joint exploitation of temperature and
polarization power spectra (Efstathiou \& Bond 1998, Kinney 1998)
is in fact required for the removal of some cosmological
parameter degeneracies.

\bigskip
We would like to thank R. Scoccimarro for useful suggestions and P. de
Bernardis, S. Masi,  and V. Natale for stimulating discussions. 
This work is partly supported by a MURST-COFIN98 (AF) and
Agenzia Spaziale Italiana-ASI (RF) grants.

\newpage

\begin{table}[t]
\begin{center}
\begin{tabular}{||c|c|c|c|c||}
\hline \hline
$C^{mn}_{\chi}$ & $z$ & $A_{max}$ & ${\Gamma}$ &
$\alpha$ \\
\hline \hline
$C^{33}_{\chi}$& 14.3 & 0.147 & $3.8\times 10^{-5}$ & 1.6 \\ \hline
$C^{44}_{\chi}$& 15.4 & 0.180 & $3.7\times 10^{-5}$ & 1.2 \\ \hline
$C^{55}_{\chi}$& 16.5 & 0.234 & $3.0\times 10^{-5}$ & 1.4 \\ \hline
$C^{66}_{\chi}$& 18.0 & 0.209 & $2.4\times 10^{-5}$ & 1.5 \\ \hline
$C^{77}_{\chi}$& 19.8 & 0.125 & $2.1\times 10^{-5}$ & 1.6 \\ \hline
$C^{88}_{\chi}$& 22.1 & 0.039 & $1.7\times 10^{-5}$ & 1.7 \\ \hline \hline

\end{tabular}
\caption{\footnotesize Autocorrelation function fitting parameters} 
\label{tab51}
\end{center}
\end{table}

\begin{figure}[b]
\centerline{\tt Figure available at
http://www.arcetri.astro.it/$\sim$ferrara/reion.html}
\caption{{\protect\footnotesize {Snapshot of the distribution of ionized
regions in the simulation box at redshift $z=19.8$ for run A}}}
\label{fig1}
\end{figure}

\begin{figure}[tbp]
\psfig{figure=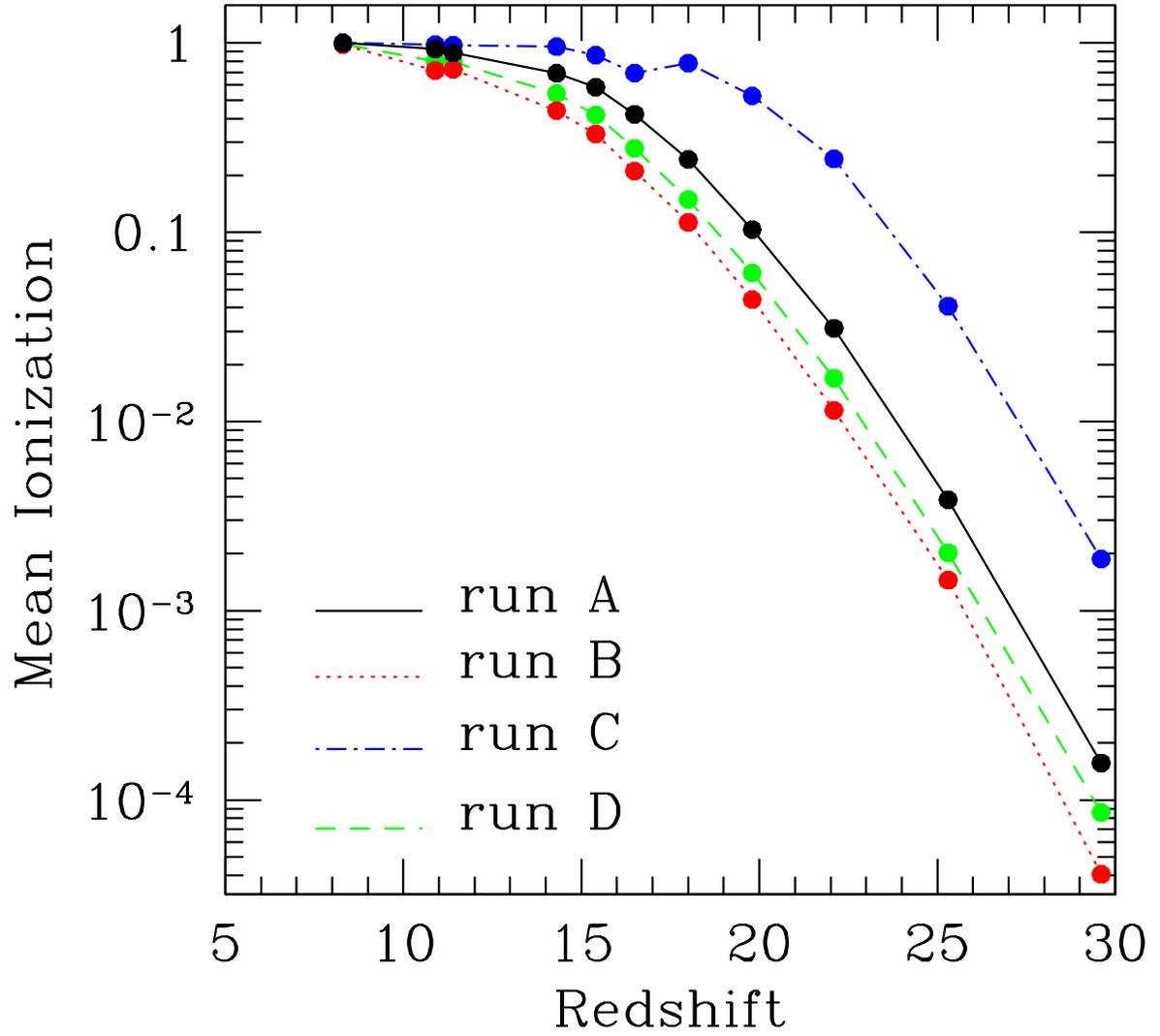}
\caption{{\protect\footnotesize {\ Evolution of the volume-averaged mean
ionization, $\bar\chi(z)$, for the four runs A, B, C, D whose parameters are given
in the text.}}}
\label{fig2}
\end{figure}

\begin{figure}[tbp]
\psfig{figure=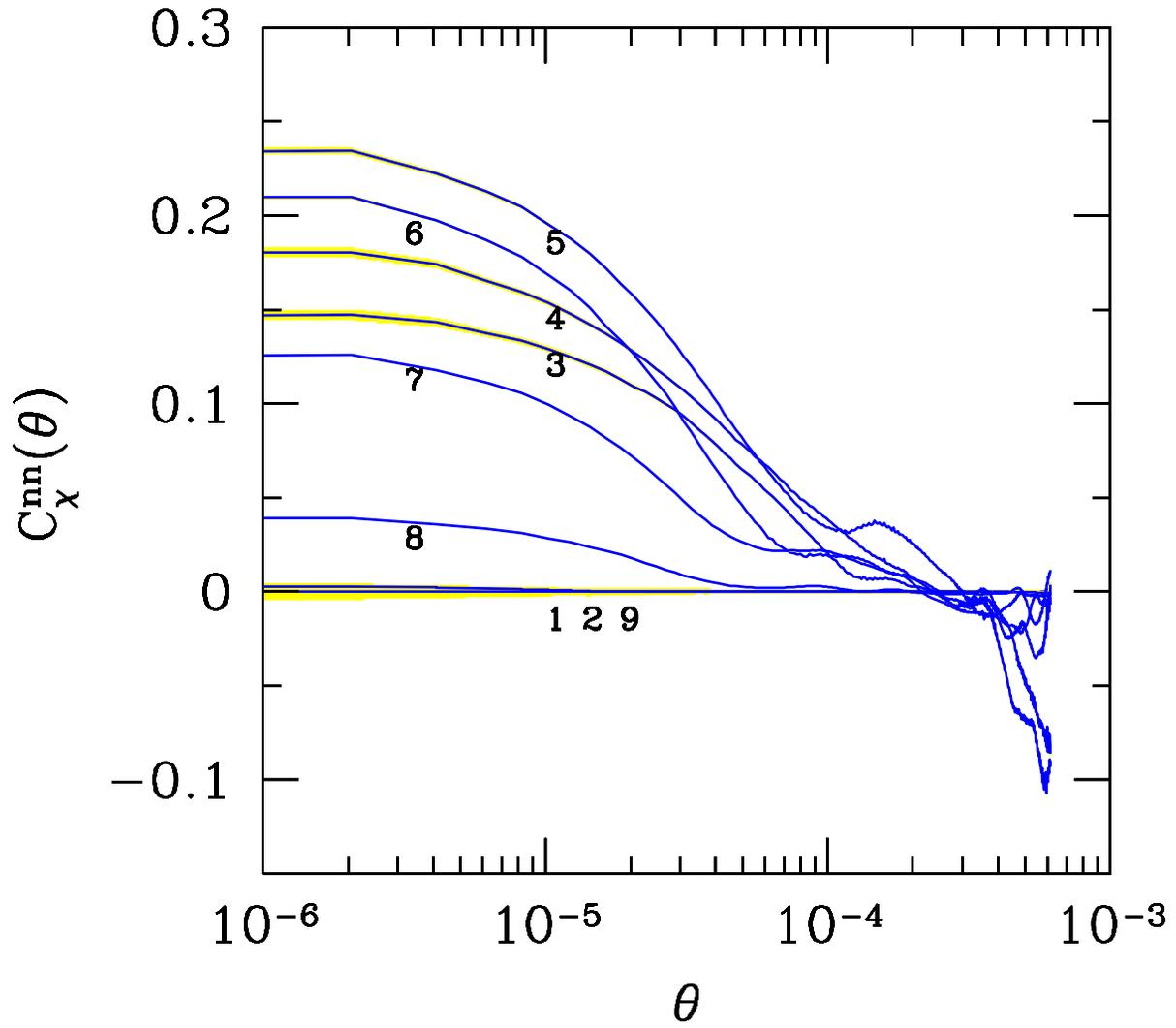}
\caption{{\protect\footnotesize {Evolution of the ionization autocorrelation
function $C_\chi^{nn}(\theta)$ with $n=1,..,9$, where $n$ labels the
redshift as explained in the text. Also shown are the statistical errors
(shaded areas).}}}
\label{fig3}
\end{figure}

\begin{figure}[tbp]
\psfig{figure=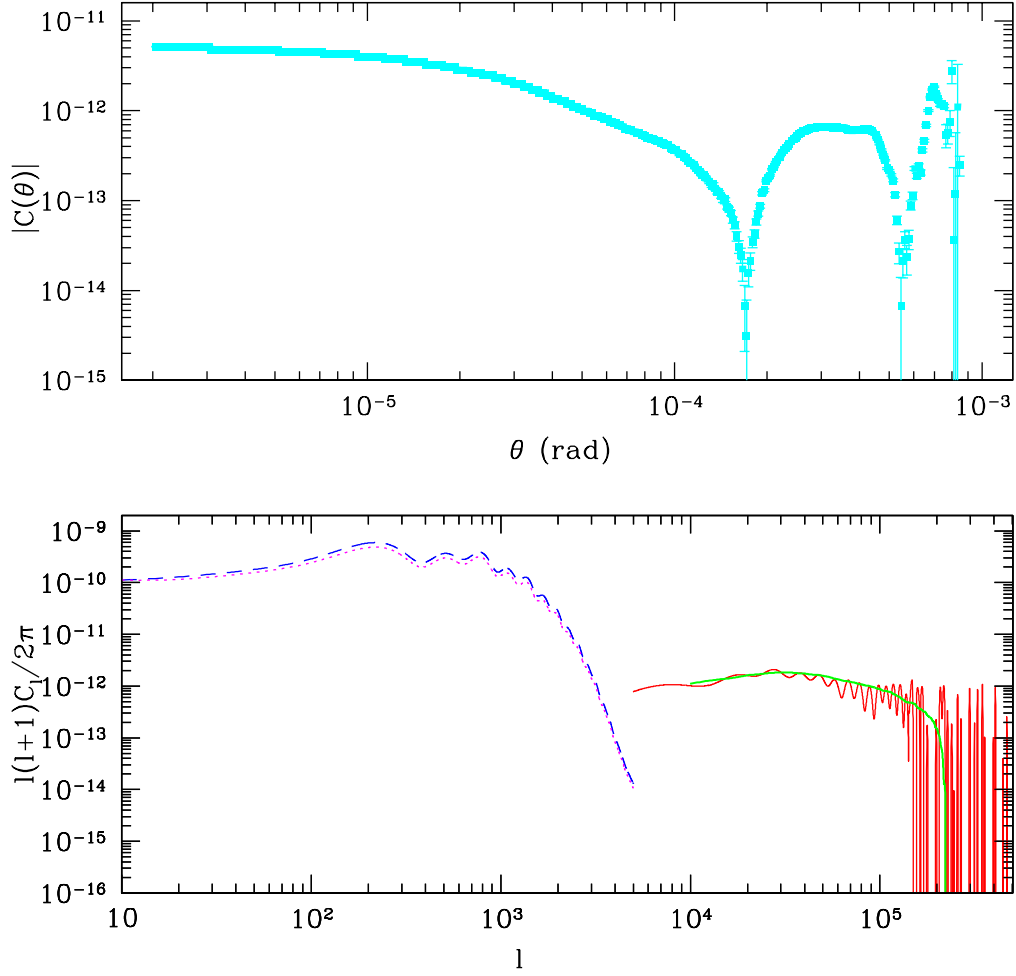, height=15cm}
\caption{{\protect\footnotesize {{\it Top:} Angular correlation function
 $C(\theta)$; {\it Bottom:} Power spectrum $
C_\ell$ of secondary anisotropies produced by IHR; superposed is the
running spectrum average $S_\ell$. The curves on the 
left are the
primary spectra in which homogenous reionization effects are neglected
({\it solid}) or included ({\it dashed}) with a reionization
optical depth $\tau =0.177$.}}}
\label{fig4}
\end{figure}

\vfill
\eject

\end{document}